\documentclass[final,english]{bullsrsl}[2022/06/15]

\usepackage[latin1]{inputenc}
\usepackage[T1]{fontenc}

\usepackage{natbib} 
\usepackage{graphicx}

\begin{document}
\title{Polarimetric Approach Towards Membership Probability of Open Star Cluster NGC 2345}

\author[affil={1,2}, corresponding]{Sadhana}{Singh}
\author[affil={2}]{Jeewan Chandra}{Pandey}
\author[affil={1}]{Vishal}{Joshi}
\affiliation[1]{Astronomy \& Astrophysics Division, Physical Research Laboratory (PRL), Ahmedabad-380009, India}
\affiliation[2]{Aryabhatta Research Institute of Observational Sciences (ARIES), Manora Peak, Nainital 263001, India}
\correspondance{sadhanasingh0824@gmail.com, sadhanasingh@prl.res.in}
\maketitle

\begin{abstract}
Using the linear polarimetric observations, we present a method to derive the membership probability of stars in cluster NGC 2345. The polarimetric observations of cluster NGC 2345 are performed using the instrument ARIES IMaging POLarimeter (AIMPOL) mounted as a backend of the 104-cm telescope of ARIES. Members of the cluster should exhibit comparable polarization since they are located nearly at the same distance. This concept is used to extract the membership probability of known member stars of cluster NGC 2345. The membership probability estimated using the polarimetric data for cluster NGC 2345 agrees with the membership probability derived from the proper motion method in the previous studies.
   
\end{abstract}


\section{Introduction}
Star clusters are the best target to probe stellar evolution, dynamics, Galactic structure, and evolution. The basic information required to study the cluster is to identify the membership of stars to the cluster. Proper motion technique has been proven as a significant way to derive membership information. However, it is possible that the proper motion technique can misguide the information of a star's membership for some peculiar stars (e.g. pre-main sequence stars, supergiants, Cepheids, etc.) \citep{1986IAUS..109..201C, 2006A&A...446..949D}.
Thus, the polarimetric technique is being explored to extract the membership information of star clusters. The polarization of starlight depends on the column density of dust grains that lie in front of the star if it is not intrinsically polarized. The aligned asymmetric dust grains present in front of the star cause the dichroic extinction of the starlight, and hence, the light gets linearly polarized  \citep{1949ApJ...109..471H, 1949Sci...109..165H,  1949Sci...109..166H}. As the star clusters are formed from the same molecular cloud, the member stars of clusters are located almost at a similar distance from the observer. Hence, the polarization study can be used to extract the membership probability of star clusters.

The polarimetric technique was found to be reliable for estimating the membership probability (MP) of known members from a proper motion study \citep{2013MNRAS.430.1334M}. Therefore, we have selected the known member stars of cluster NGC 2345 to derive the MP using the polarimetric technique.  The basic parameters of the cluster NGC 2345 compiled from \citet[]{2005A&A...438.1163K, 2013A&A...558A..53K, 2018A&A...618A..93C, 2020A&A...633A..99C}, are given in Table  \ref{tab:clus_param}. 

\begin{table}[h]
\centering
\begin{minipage}{138mm}
\caption{Basic parameters of the open star cluster NGC 2345.}
\label{tab:clus_param}
\end{minipage}
\bigskip
\begin{tabular}{p{0.4\textwidth}|p{0.3\textwidth}}
\hline
\textbf{Cluster parameters} &  \textbf{NGC 2345} \\
\hline
Longitude (degree) & 226.58  \\
Latitude (degree) & -2.31  \\
Age (Myr) & 55-79  \\
Distance (kpc) & 2.2-3.0 \\
Reddening [$E(B-V)$] (mag) & 0.59-0.68  \\
Core radius (arcmin) & 4  \\
Cluster radius (arcmin) & 7  \\
\hline
\end{tabular}
\end{table}

\section{Method}
\subsection{Polarimetric Data}
The polarization values in V-band for the open cluster NGC 2345 are taken from our previous study \citep{2022MNRAS.513.4899S}. The cluster was observed from the 104-cm telescope of ARIES using the AIMPOL instrument \citep{2004BASI...32..159R,2023JAI....1240008P} on 22 January, 14 and 15 February 2018. The observations were performed at four different positions of the half-wave plate, i.e., 0, 22.5, 45, and 67.5 degrees from the celestial north-south direction. The detector was a 1k$\times$1k Charge-Coupled Device camera (CCD) having read noise of 7.0 $e^{-}$ and gain of 11.98 $e^{-}/ADU$. Further observation details of this cluster can be found in \citet{2022MNRAS.513.4899S}. After cleaning images, we performed aperture photometry to get the fluxes of ordinary and extraordinary images using the Image Reduction and Analysis Facility. The detailed data reduction procedure to extract the degree and angle of polarization is described in \citet{2004BASI...32..159R, 2020AJ....159...99S, 2020AJ....160..256S}. Correction for instrumental polarization and zero-point polarization angle was performed using the unpolarized standard star (HD 21447) and polarized standard stars (HD 19820 and HD 25443). We have used the $V-$ band polarimetric data to derive Stokes parameters $Q$ and $U$ using $Q = P cos2\theta$ and $U = P sin2\theta$, where $P$ is the degree of polarization and $\theta$ is the position angle. The membership information based on the proper motion for polarimetrically observed stars in the cluster is taken from \citet{2018A&A...618A..93C}. We have considered only those stars as members of the cluster with the MP $\geq$ 50 \% as the distribution of MP of member stars peaks between 50 to 100\%.

\subsection{Estimation of Membership Probability}
\begin{figure}[h]
\centering
\includegraphics[scale=0.6]{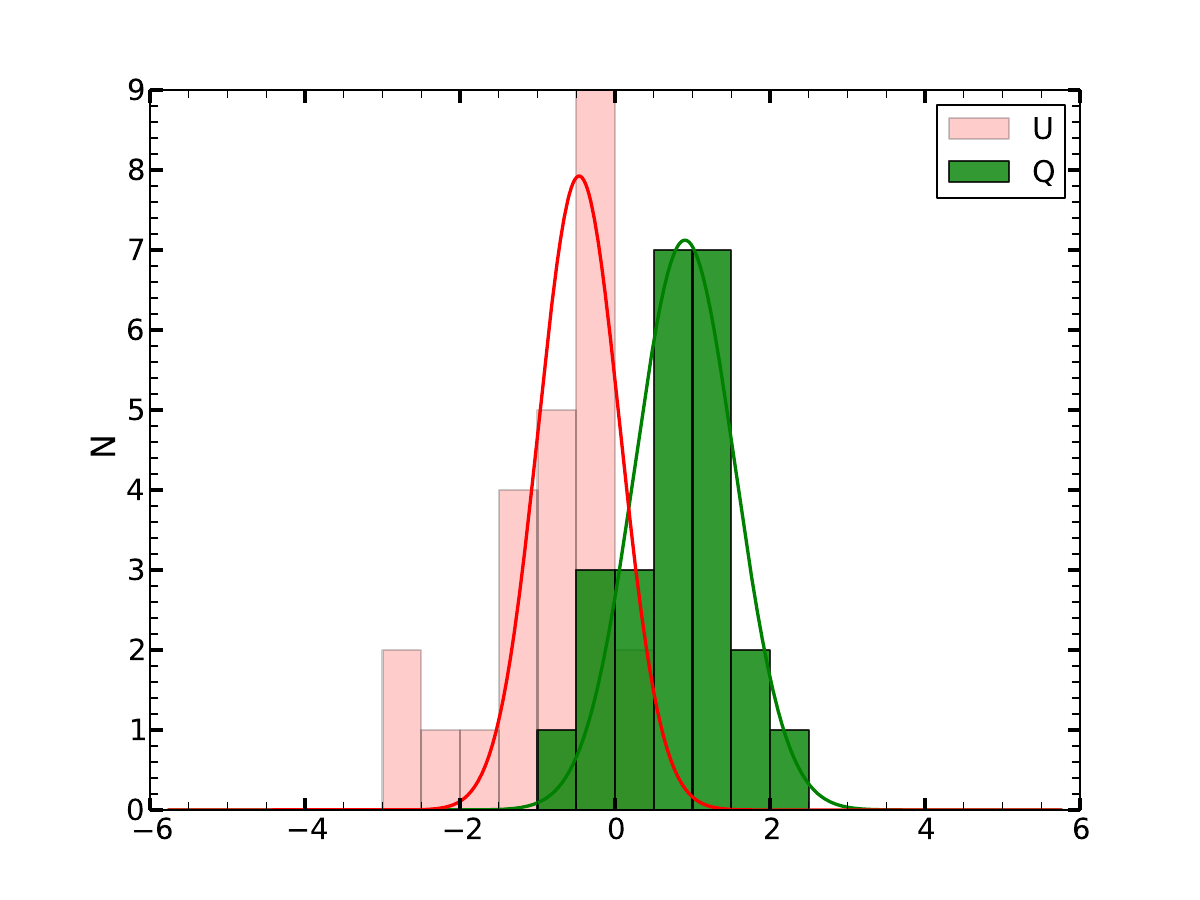}
\bigskip
\begin{minipage}{14cm}
\caption{The distribution of Stokes parameters ($Q$ and $U$) among stars with MP$\geq$90\% of the cluster NGC 2345. The green and red curves show the Gaussian fit of the  $Q$ and $U$ distribution, respectively.} 
\label{fig:qu_hist}
\end{minipage}
\end{figure}

We have chosen a group of stars with very high MP for the best representation of the cluster. Therefore, we have taken stars with MP $\geq$ 90\%  for estimating the cluster's average Stokes parameters. The distribution of the cluster's Stokes parameters is shown in Figure \ref{fig:qu_hist}. The mean value of Stokes parameters for the cluster is derived by fitting the Gaussian curve to the distribution and found as $Q_{mean}=0.9026 \%$, $U_{mean}=-0.4624 \%$ for NGC 2345.  

Next, the average scatters of Stokes parameters for a star are calculated from the mean values of the cluster's Stokes parameters ($Q_{mean}$, $U_{mean}$) as 
\begin{equation}
    scatt = \sqrt{(\Delta Q)^{2} + (\Delta U)^{2}}
    \label{eq:scatt}
\end{equation}
where $\Delta Q = Q - Q_{mean}$ and $\Delta U = U - U_{mean}$. 
Additionally, the full range of Stokes parameters considering all observed stars in the cluster is calculated using the equation as 
\begin{equation}
    range = \sqrt{Q_{range}^2 + U_{range}^2}  
    \label{eq:range}
\end{equation}
where $Q_{range} = (Q_{max} - Q_{min})$ and $U_{range} = (U_{max} - U_{min})$.

Then, by comparing the average scatters with the full ranges and with the 100\% normalization, we have estimated MP as  
\begin{equation}
    MP = 100 \times \left(\frac{range - scatt}{range}\right)
    \label{eq:mp}
\end{equation}

The estimated MP of member stars for the cluster NGC 2345 are given in Table \ref{tab:mp_mem} along with the MP's of these stars from the literature. The first column is the star ID taken from \citet{2022MNRAS.513.4899S}. 

\begin{table}[h!]
\centering
\begin{minipage}{14cm}
\caption{MP of member stars for the cluster NGC 2345 using the polarimetric technique along with MP of these stars from other studies.}
\label{tab:mp_mem}
\end{minipage}
\bigskip
\footnotesize	
	\begin{tabular}{cccc}
	\hline
	\multicolumn{1}{c}{\textbf{ID}} & \multicolumn{3}{c}{\textbf{Membership Probability}}  \\
	\hline
		\multicolumn{1}{c}{} & \multicolumn{1}{c}{This work}  & \multicolumn{1}{c}{ \citet{2018A&A...618A..93C}} & \multicolumn{1}{c}{ \citet{2017MNRAS.470.3937S}}\\
	\hline
6     &  0.75   &    0.9  &	0.91  \\
14    &  0.89   &    0.9  &	0.93  \\	
17    &  0.90   &    0.9  &	0.85  \\	
25    &  0.81   &    0.9  &	0.0   \\	
28    &  0.94   &    1.0  &	0.90  \\	
29    &  0.91   &    0.9  &	0.93  \\	
32    &  0.90   &    0.9  &	0.92  \\	
34    &  0.76   &    0.8  &	0.93  \\	
37    &  0.99   &    1.0  &	0.93  \\	
38    &  0.86   &    1.0  &	0.89  \\	
39    &  0.96   &    1.0  &	0.93  \\	
41    &  0.98   &    1.0  &	0.93  \\	
42    &  0.97   &    1.0  &	0.93  \\	
43    &  0.95   &    1.0  &	0.93  \\	
46    &  0.93   &    1.0  &	-     \\	
48    &  0.77   &    1.0  &	0.74  \\	
49    &  0.58   &    0.7  &	0.35  \\	
50    &  0.93   &    1.0  &	0.91  \\	
51    &  0.85   &    0.7  &	-     \\	
58    &  0.93   &    1.0  &	-     \\	
64    &  0.97   &    0.7  &	-     \\	
66    &  0.88   &    0.8  &	0.93  \\	
68    &  0.83   &    1.0  &	0.93  \\	
69    &  0.91   &    0.9  &	0.93  \\	
72    &  0.94   &    1.0  &	0.91  \\	
82    &  0.93   &    1.0  &	0.89  \\	
84    &  0.97   &    1.0  &	0.93  \\	
90    &  0.97   &    0.9  &	-     \\	
91    &  0.97   &    1.0  &	0.93  \\
\hline
\end{tabular}
\end{table}

\section{Discussion and Conclusions}
\begin{figure}[h]
\centering
\begin{tabular}{cc}
\includegraphics[scale=0.3]{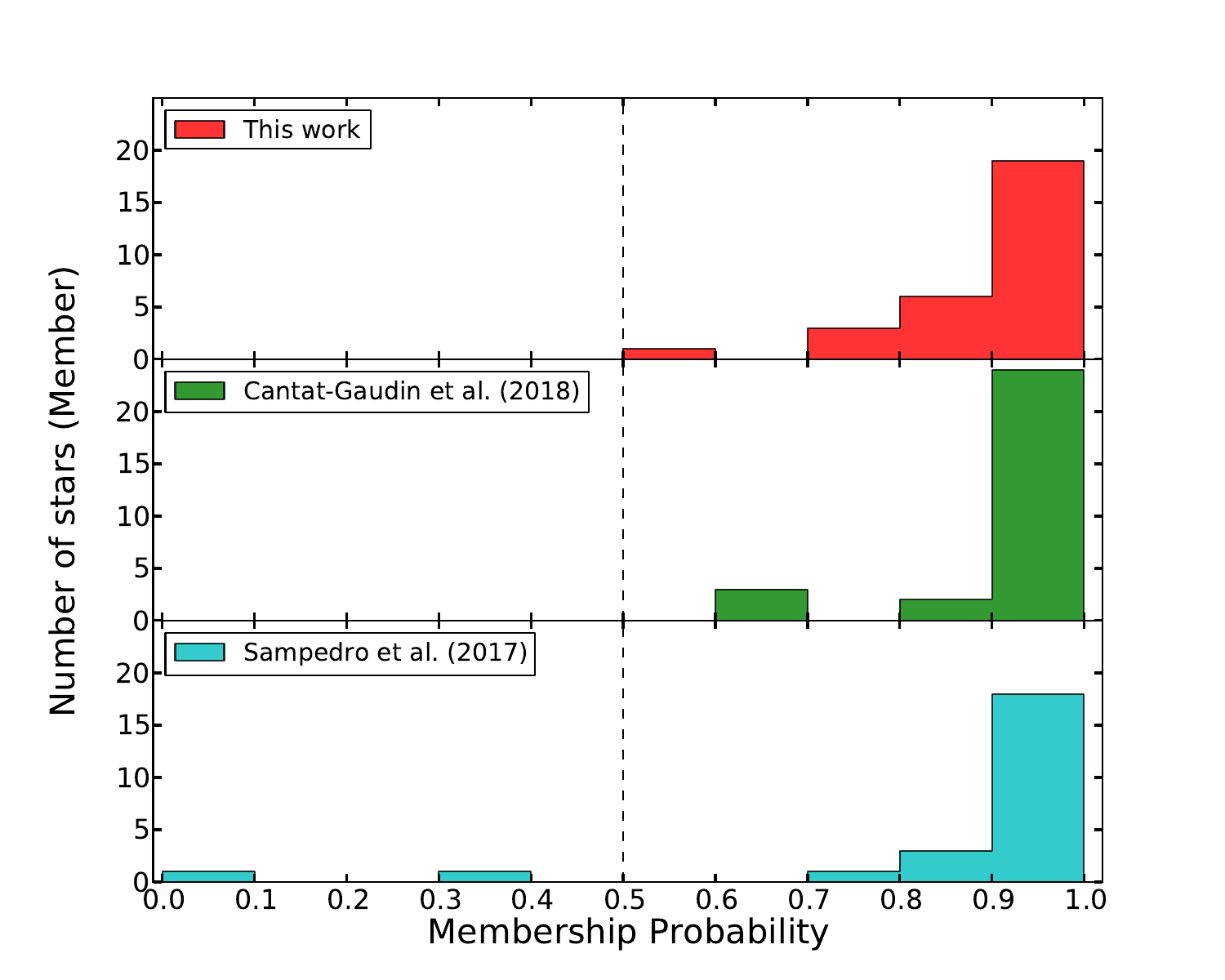} &
\includegraphics[scale=0.3]{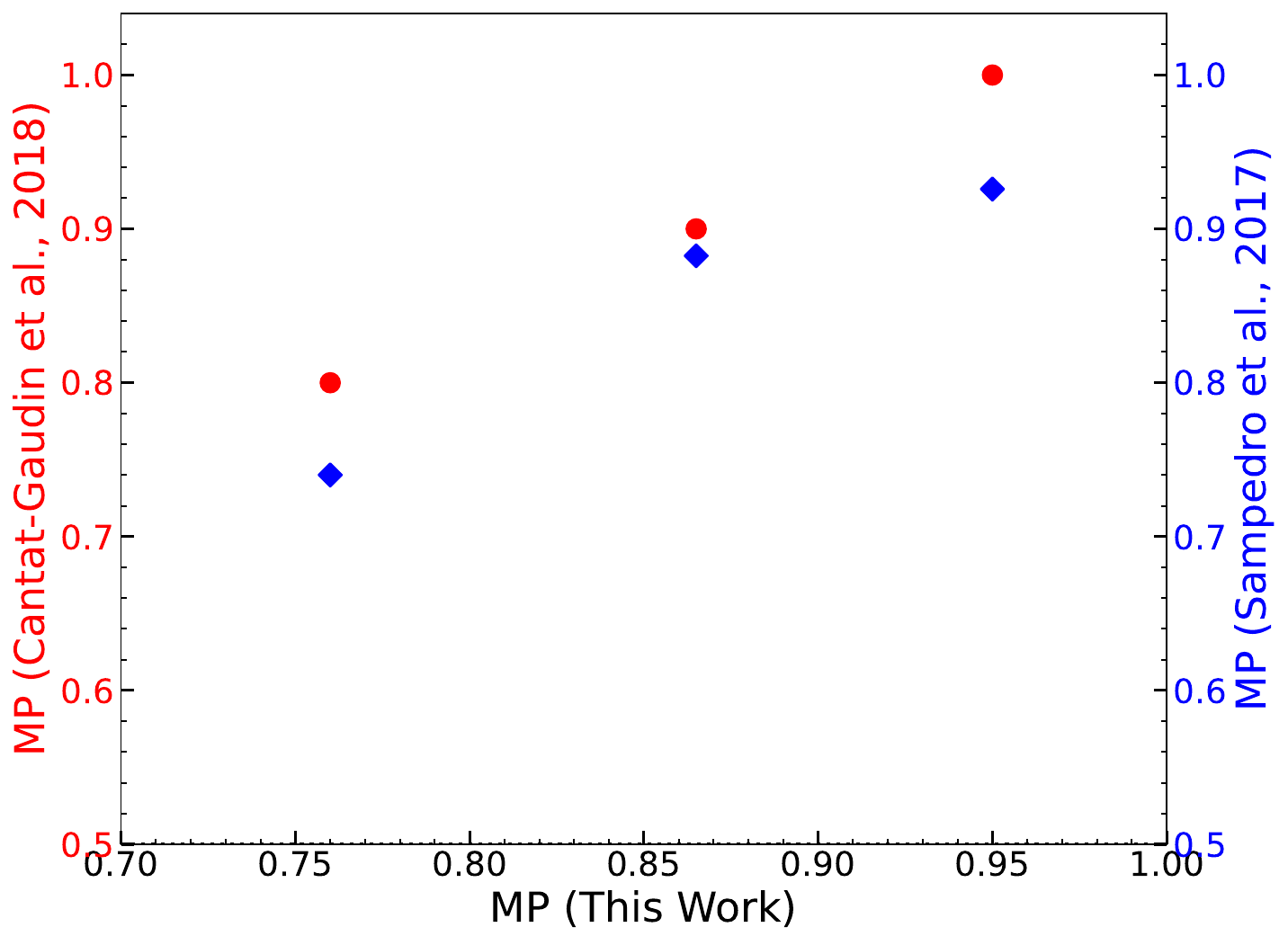} 
\\
(a) & (b) \\
\end{tabular}
\bigskip
\begin{minipage}{14cm}
\caption{Comparison of MPs of proper motion member stars of the open star cluster NGC 2345 in different studies. (a) The vertical line at 0.5 denotes the MP of 50\%. (b) The red scale on the left side shows the MP from  \citet{2018A&A...618A..93C}, and the blue scale is for MP from \citet{2017MNRAS.470.3937S}. The markers denote the average binned data points of bin size 0.1 in MP.}
\label{fig:2345_mp}
\end{minipage}
\end{figure}

Using the polarimetric technique, we have extracted membership probabilities of previously derived member stars of cluster NGC 2345. A plot for comparing our results of polarimetric MP with the proper motion MP reported in previous studies is shown in Figure \ref{fig:2345_mp}, and a good agreement between our results and previously estimated probabilities of members is found. 

\begin{furtherinformation}
\begin{orcids}
\orcid{0009-0000-2098-6119}{Sadhana}{Singh}
\orcid{0000-0002-4331-1867}{Jeewan Chandra}{Pandey}
\orcid{0000-0002-1457-4027}{Vishal}{Joshi}
\end{orcids}

\begin{authorcontributions}
All authors contributed significantly to the work presented in this paper.
\end{authorcontributions}

\begin{conflictsofinterest}
The authors declare no conflict of interest.
\end{conflictsofinterest}

\end{furtherinformation}

\bibliographystyle{bullsrsl-en}

\bibliography{ref_bina}

\end{document}